\documentclass[sigconf]{acmart} 
\usepackage{amsmath}
\usepackage{amssymb}
\usepackage{subfig}
\usepackage{graphicx}
\usepackage{color,url}
\usepackage[subtle,margins=normal]{savetrees}
\usepackage[]{algorithm2e} 
\usepackage[titletoc]{appendix}
\usepackage{enumitem} 

\def\BibTeX{{\rm B\kern-.05em{\sc i\kern-.025em b}\kern-.08emT\kern-.1667em\lower.7ex\hbox{E}\kern-.125emX}}

\newcommand{\paragraphb}[1]{\vspace{0.03in}\noindent{\bf #1} }

\setcopyright{none}
\renewcommand\footnotetextcopyrightpermission[1]{} 
\begin{document}

\title{Clicktok: Click Fraud Detection using Traffic Analysis}

\author{Shishir Nagaraja, Ryan Shah}
\email{{shishir.nagaraja, ryan.shah}@strath.ac.uk}
\affiliation{%
  \institution{University of Strathclyde}
}

\begin{abstract}
  Advertising is a primary means for revenue generation for millions
  of websites and smartphone apps. Naturally, a fraction abuse
  ad networks to systematically defraud advertisers of their
  money. Modern defences have matured to overcome some
  forms of click fraud but measurement studies have reported that a
  third of clicks supplied by ad networks could be
  clickspam. Our work develops novel inference techniques
  which can isolate click fraud attacks using their
  fundamental properties. We propose two defences, {\em
      mimicry} and {\em bait-click}, which provide clickspam detection
    with substantially improved results over current
    approaches. Mimicry leverages the observation that organic
    clickfraud involves the reuse of legitimate click traffic, and
    thus isolates clickspam by detecting patterns of click reuse
    within ad network clickstreams. The bait-click defence leverages
    the vantage point of an ad network to inject a pattern of bait
    clicks into a user's device. Any organic clickspam generated involving the bait clicks will be subsequently recognisable by the ad network.  Our
  experiments show that the mimicry defence detects around 81\% of
  fake clicks in stealthy (low rate) attacks, with a false-positive
  rate of $110$ per hundred thousand clicks. Similarly, the
    bait-click defence enables further improvements in
  detection, with rates of ~95\% and a reduction
  in false-positive rates of between 0 and 30 clicks per million
  -- a substantial improvement over current approaches.
\end{abstract}

\maketitle




\section{Introduction}



By definition, click fraud generates no revenue for the
advertiser, but inflicts losses on tens of thousands of online
advertisers in the order of hundreds of millions of dollars each
year~\cite{mungamuru:fc08,zeroaccess}. Typically, click
fraud is generated by malicious applications
  (apps) and malware, and is responsible for around 30\%
of click traffic in ad
networks~\cite{dave:sigcomm:2012,bluff}.

Early threshold-based defences demonstrated a focus on the
volumes of click fraud from bad-listed sources, but failed
  simply because attackers were able to swiftly discard
their publisher accounts after receiving
  reputational hits from ad network
defences~\cite{zhang:webquality:2011,dave:ccs:2013}, and opened new
ones. The decrease in the cost of botnet rentals in the
  underground economy has been a primary driver of
  fraud~\cite{botrent09, wei2016ddos, miraicost}. Click
  botnets can generate massive amounts of fake traffic and ad
  impressions, by automatically clicking on websites and apps in large
  numbers. The reduction in the fixed costs of generating
  clickspam, by several orders of magnitude, significantly
  reduces the number of fake clicks per host required to run an
economically sustainable click fraud operation. Assuming the
  earning potential of 0.5 cents per click, which is at the lower end
of the spectrum, an attacker can cover operational costs
with just three to five fake clicks a day, with a few more to run
their operation profitably. Based on this, traditional
threshold-based
defences~\cite{dave:ccs:2013,dave:sigcomm:2012,metwally:www07,metwally:vldb08}
fail as the levels of click fraud per source goes below the
detection threshold.

Subsequently, ad network defences have evolved to generate
liveness-proofs by running machine learning models on advertisement
clickstreams to distinguish bots from humans. As a
response, attackers mimic the actions of legitimate device
users in order to generate credible click
fraud~\cite{zeroaccess,pubcrawl}. Recently, a serious increase in
organic click fraud has been noted via mobile
malware. Fraudsters develop seemingly legitimate apps or
  purchase those with high reputation scores. These apps
  carry out a legitimate activity, such as controlling the torch, but
  also serve as a mechanism for mining the (organic) click
    activity of the device user. Moreover, attackers
  then launder mined clicks back through their
  installed user-base. Since the click fraud is based on legitimate
  traces, the clicks are able to pass through ad network
  filters. The exception is where the attack violates a threshold,
  such as using a small pool of IP addresses to carry out the
  attack. Ultimately, this motivates the need for automated
  detection techniques that can scale the detection of click fraud
  attacks, whilst ensuring the integrity of the digital
  advertising ecosystem.

\begin{figure}[!ht]
\begin{center}
\leavevmode
\includegraphics[width=1\columnwidth]{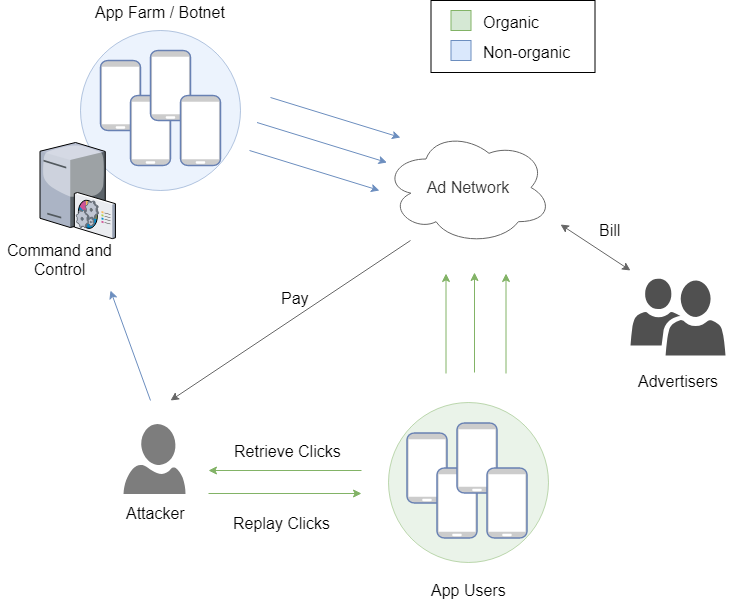}
\end{center}
\caption{Organic and Non-Organic Click fraud}
\label{fig:onofraud}
\end{figure}

To address this, we developed Clicktok, a statistical technique that
detects clickspam by efficiently searching for {\em
    reflections} of click traffic, encountered by an ad network in
the past. Clicktok is based on exploiting the timing properties
of click traffic, and also unifies the technical
response, offering a defence technique which isolates both
organic, and the relatively simpler, non-organic click fraud
attacks. The focus of this work is on generic click fraud defences,
rather than analysing individual click-modules. Current
  efforts to defeat click fraud have primarily focused on fraud
  measurement techniques~\cite{dave:sigcomm:2012}, measurement and
  analysis of publisher fraud~\cite{dave:ccs:2013,chachra:imc:2015},
  and ad placement fraud~\cite{DECAF}. While publisher and affiliate-
  marketing fraud are doubtless of importance, there is limited work
  that focuses on detecting click fraud from ad network clickstreams
  supplied to advertisers. Upon evaluation of our defence, we report
  significantly improved detection and estimation over past efforts
   ~\cite{dave:sigcomm:2012,dave:ccs:2013,
    metwally:www07,metwally:vldb08, metwally:icdcs07}.

\paragraphb{Contributions:} First, we have developed two clickspam
defenses: the {\em mimicry defence} and the {\em bait-click
  defence}. Second, we have developed a unified algorithm for
detecting both organic and non-organic clickspam, which
provides the engineering advantage that ad networks need
only to adopt a single defence. Finally, our techniques
support attribution via their ability to separate malware
clicks embedded within legitimate clickstreams. This
  demonstrates the usefulness of the algorithm in building passive
and active click fraud defences using real-world data.

\paragraphb{Roadmap:} We start by giving a detailed problem
description in Section~\ref{sec:problem}. In Section~\ref{sec:inference}, we
describe our overall approach and algorithm. We then evaluate the
performance of our algorithm on click fraud traffic, embedded within
real click traffic data in Section~\ref{sec:eval}. Related work is
situated at the end of the paper.

\section{The click fraud detection problem}
\label{sec:problem}
Click fraud involves directing fraud clicks (or clickspam) at online
advertisements (ads), concerning three parties: an
advertiser, a publisher, and an ad network. The advertisers
participate in a keyword auction, organized by the ad
  network, where ads owned by the winning advertiser are submitted for
  circulation by the ad network. The publisher's role is to
  render the advertisements that are provided by the ad network. When
a user clicks on an ad, the ad network: receives the
request, updates the billing account of the corresponding advertiser,
and redirects the click to a URL of the advertiser's choice. For each
click on an ad, the advertiser pays the ad network, who
in turn pays the publisher a substantial fraction of the per-click
revenue ($\approxeq 70\%$).

\begin{figure}[!ht]
\begin{center}
\leavevmode
\includegraphics[width=\columnwidth]{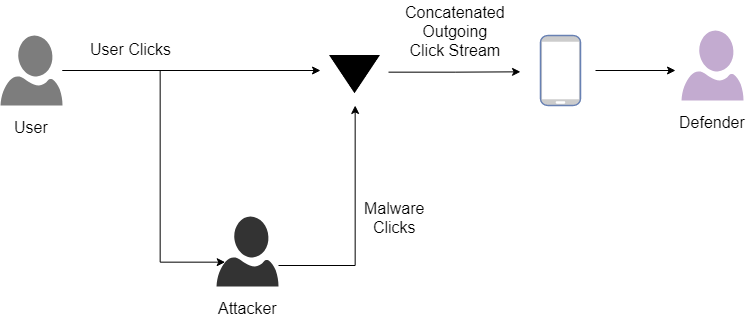}
\end{center}
\caption{Click fraud detection problem}
\label{fig:problem}
\end{figure}

The click fraud problem is the challenge of distinguishing
  between clickspam and legitimate clicks, given that the attacker has
  full knowledge of legitimate click traffic
  (Figure~\ref{fig:problem}). Typically, an attacker
  registers as a (large number of) publishers with the ad
  network. They deploy techniques to send clickspam to
advertisements, allowing them to gain a fraction of the
money paid out by the advertiser for every (fake) click.

\paragraphb{Terminology:} A paper on click fraud makes
  frequent use of certain terminology, which we now cover for ease of
  readability. {\em Click traffic}, {\em click stream}, or {\em click
    traces} all refer to an ordered sequence of one or more clicks
  each corresponding to an advertisement. Clicks are termed {\em
    legitimate} or {\em organic} when generated by a genuine
  user. Clicks generated for the motive of profit are considered {\em
    fake}, {\em fraudulent}, or {\em clickspam}. Clickspam generated
  using legitimate traces is termed {\em organic} clickspam.

\subsection{Challenges}
\label{sec:attack}


Efficient clickspam detection involves several
  challenges. First, an ideal detection system should be able to
directly observe human input on end user devices, and label click
traffic that is suspected to be illegitimate; however this
is an impractical approach. Second, legitimate
click traffic that arrives at an ad network
likely dwarfs the {\em relatively} smaller amount of
clickspam. Third, the variable form and structure of click
  traffic poses an issue when constructing an accurate baseline model
of legitimate click traffic, that is broadly
applicable. Finally, click malware can employ a
  variety of stealth techniques to evade detection, in particular the
use of organic click traffic and reducing the number of fake clicks
per source, to well below the level set by threshold-based
click fraud detection techniques.

\subsection{Opportunities}
\label{sec:opp}
Although adapting fake clicks to match the statistical
characteristics of legitimate clickstreams is undoubtedly a stealthy
approach, we note that this can be used as a point of
detection. Based on this, we identify both passive and
  active approaches to defence.

\paragraphb{Passive Approach:} We argue that legitimate
click activity has copy-resistance properties, owing to the
uncertainty of inter-click times. Therefore, to accurately
mimic a user's click activity, an attacker
would need to model the timing behaviour of the
  user, which has some uncertainty that forms the basis
for our passive defences. Furthermore, click
generation techniques, that are encoded by malware authors,
may result in correlations in the inter-click times of clickstreams
across users. Thus, another approach is to consider the
relative increase in the correlation across clickstreams,
due to click fraud.

\paragraphb{Active Approach:} While considering a passive
  approach to defending against click fraud, a radically different
  approach is to consider active interventions. By adding
           {\bf bait clicks} to legitimate user traffic, we
             can attract the attention of malware. To
             measure and understand malware click generation
             strategies, malware adaptation to a bait clickstream may
           be a promising approach. This is a novel idea for
           instrumenting click malware --- the ad network injects a
           pattern of clicks into a user device over a period of time,
           via a client-side scripts executed by the browser. Only
           malware will respond to this click pattern,
           mistaking it for legitimate activity, whilst the ad network
           ignores these clicks. If click malware is to be present on
           the device and adapts to the user's activity, it
           will generate clickspam that can be readily isolated.


\section{Inference System}
\label{sec:inference}

\subsection{System Architecture}

\begin{figure}[!ht]
\begin{center}
\leavevmode
\includegraphics[width=\columnwidth]{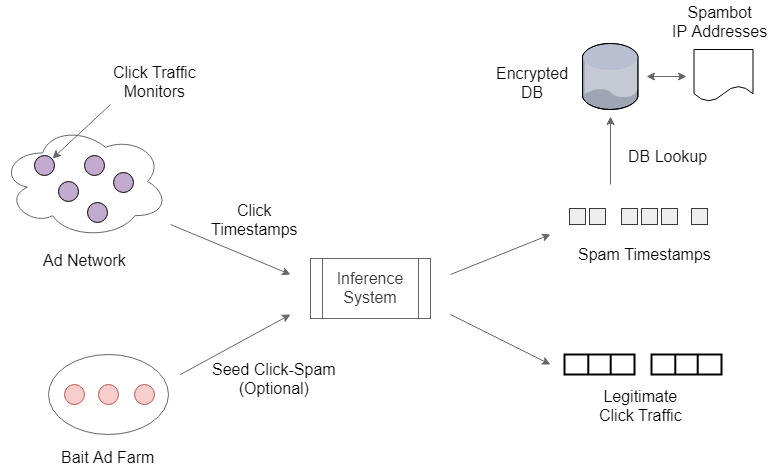}
\caption{System Architecture}
\label{fig:arch}
\end{center}
\end{figure}

We propose an inference system, depicted in Figure~\ref{fig:arch}, which
takes two inputs, click timestamps from the ad network, and an optional
seed clickspam input from a bait ad farm. The ad network contains
servers which run {\em click traffic monitors} that store click timestamps.
Optionally, to supplement information for click traffic monitors, our
inference system may also receive input from a bait ad farm (honeynet)
~\cite{bluff}. Classification information from these may be used as
an input to Clicktok, where the suspicion of click fraud is known
with a higher probability.

Our primary focus is to design a generic inference
algorithm that is, first, based on the fundamental
limitations of automated fake click generation techniques, and second,
can address both organic and non-organic clickspam.

Clicktok works on the core observation that both organic and inorganic
clickspam cause an increase in redundancy, albeit differently within
ad network clickstreams. In the case for organic click
  fraud, to isolate the source of redundancy, we use a compression
function in combination with a clustering algorithm, to isolate click
traffic whose {\em timing patterns} are similar to past timing
patterns. A timing pattern is an ordered ascending sequence of time
offsets, relative to an absolute start time.
  
  Similarly, we noticed that the same intuition can be
  leveraged to isolate inorganic clickspam. For instance, where
  malware generates traffic using randomised generators, the traffic
  with high entropy timing patterns can be clustered together, by
  exploiting their non-compressibility. Likewise, the
  injection of small amounts of clickspam per device are evident, when
  traffic from multiple end-user devices is
  considered together, thus exploiting the
  common patterns across infected devices. Ad networks, or backbone
  routers, provide us with a vantage point into  click traffic
  tainted with clickspam.


The primary challenge in partitioning on the basis of inter-click
times, is that both clickspam and legitimate clicks may not
be temporally separated. As Figure~\ref{fig:inputpattern} be superimposed over each
other. Interestingly, click modules, such as
Zeroaccess~\cite{zeroaccess} and its variants, have been documented to
distribute clickspam {\em only} after detecting
some legitimate activity on a device. As well as this,
another behaviour has been observed, which involves
attackers adding random time offsets to blend in.

Overall, Clicktok addresses the following challenges:

\begin{enumerate}[label=(\alph*)]
	\item{
		Resisting a mimicking source --- an adversarial source that imitates a
		partially observed source, and
	}
	\item Superimposed legitimate and clickspam time series.
\end{enumerate}

\subsection{Inference Algorithm}
\label{sec:inference:algdesign}

\begin{figure}[!htbp]
\begin{center}
 \includegraphics[width=1\linewidth]{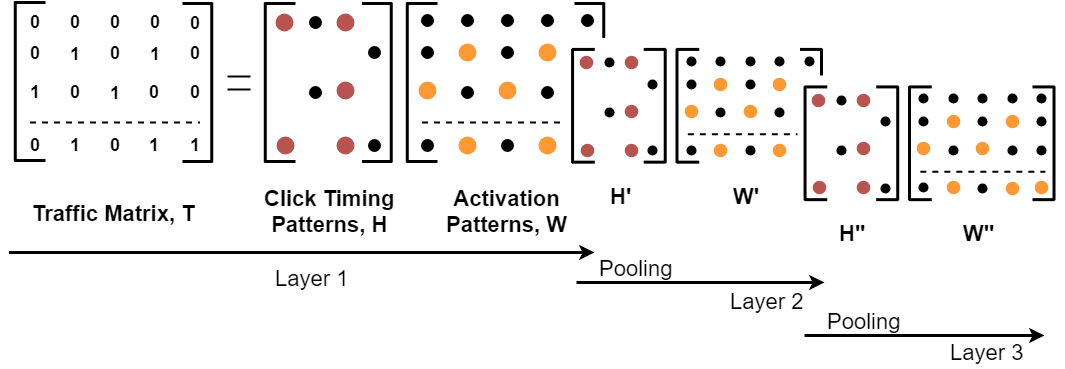}
\end{center}
\caption{Multi-layer NMF}
\label{fig:alg}
\end{figure}


Clicktok uses a decomposition technique (Figure~\ref{fig:alg}), which
partitions click traces into legitimate clicks and clickspam. We make
no assumptions about the shape or form of legitimate traffic, however
we assume that click fraud is a minority fraction (< 50\%) of the
traffic traces.  Partitioning leverages the observation that the shape
and form of organic clickspam has a close dependency on the
legitimate trace that was used to generate a specific clickspam
attack. Partitioning is carried out using a multi-layer non-negative
matrix factorization (NMF) algorithm.

\paragraphb{Traffic matrix construction:} Traffic traces from $n$
source identifiers (e.g. source IP addresses or Android ID) are
collected at a suitable vantage point such as an ad network or
enterprise backbone, to construct a $n$x$m$ traffic matrix of
observations $O$. Each element $O_{ij}$, contains the number of clicks
from source $i$ during time-interval $j$.

The goal of the inference algorithm is to decompose the input
clickstream into constituent $r$ highly sparse timing patterns called
basic-patterns. A basic pattern $h_i$ can be represented as:
$h_i={c+m, \dots, c+k}$ where $c$ is the start time, and \(0 < m <
\dots < k < \infty\) are offsets. Inference is achieved by the notion
of compression. The idea is that the input traffic traces can be
compressed down into $r$ basic-patterns and weights. Specifically, the
traffic matrix $O$ is decomposed into $H$ and $W$ as:

\begin{equation}
 O = HW
\label{eqn:nmf}
\end{equation}

\begin{figure}[!ht]
\begin{center}
\leavevmode
\includegraphics[width=0.7\columnwidth]{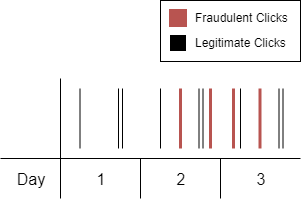}
\end{center}
\caption{Overlapping legitimate and fraud clicks}
\label{fig:inputpattern}
\end{figure}

While using compression to trace stolen click traffic and their usage
in click fraud campaigns has obvious potential, the main technical
challenge is that click fraud campaigns utilise legitimate traffic as
cover as shown in Figure~\ref{fig:inputpattern}. This creates
interleaved (superimposed) click traffic which must be unmixed with
minimal assumptions about baseline user or attacker behaviour. If this
were not the case, a simple application of time-series correlation
analysis~\cite{wei:1994:time} would reveal click fraud.

The inference algorithm illustrated in Figure~\ref{fig:alg} has two
steps, first multiple layers of partitioning and second a pooling
step. Both steps are motivated by Deep Neural Networks
(DNN)~\cite{alg:deep} and are prior art.

\paragraphb{Step1: Traffic partitioning} First, nested
  layers of NMF algorithm~\cite{nips00:leeseung} partition the traffic
  matrix $O$. NMF partitions the observed click matrix into sparse
  timing patterns (matrix $H$) and activation patterns (matrix $W$),
  i.e. $O=HW$. Sparsity is a key property here that incorporates the
  intuition of compressive partitioning within the decomposition
  step. NMF is an iterative technique with a multiplicative
  optimisation function at its core and a stopping critieria of
  $||O-HW|| \leq \epsilon$ where $||\odot||$ is the Frobenius
  norm. Nested decomposition layers further promote sparsity.  Thus
  the output of the final decomposition layer is \( O^K = H^1 (H^2
  (H^3(....H^KW^K))))\), and the optimisation function works out to:
  \( \min \mid\mid O^1 - H^1 (H^2 (H^3 ( \dots (H^K W^K))\dots)
  \mid\mid \)

\paragraphb{Step2: Pooling.} Second, to reduce sensitivity
  to synchronisation errors arising from timing misalignment, a moving
  window function is applied. In DNN literature~\cite{alg:deep}, this
  is termed as a pooling function. Without this, time-synchronisation
  errors can cause redundant timing patterns that are slightly
  time-shifted, to appear within $H$. Pooling can provide some
  robustness against such errors. We incorporate pooling by including
  average-pooling~\cite{alg:pooling} in each partitioning step. This
  is a moving-window function $F$ whose input is the rows of the
  weight matrix $W^k$, and its output is the average over the input
  window $(j-c,j+c)$. \(F(W^k_{i*}) = \overline{ \{W^k_{ip}}\ |\ j-c
  \leq p \leq j+c, 0 \leq j \leq N\). To absorb alignment errors of up
  to an hour, we set $c=12$ in our experiments.


\begin{algorithm}
    \KwIn{
        \(O \in \mathbb{R}^{m \times n}\)\\
        The number of layers K\\
        The number of columns to pool across, poolsize}
    \KwOut{Partitioning at each layer $k$}
     \For{$k$ in $1:K-1$}
         {
           \While{$\epsilon>0.05$}
            {
             \(H^k = H^k \odot \frac{O^k (W^k)^T}{H^k W^k (W^k)^T}\)\\
             \(W^k = W^k \odot \frac{(H^k)^T O^k}{(H^k)^T H^k W^k}\)\\

             \(\epsilon = \sqrt{\sum_i^m \sum_j^n (O^k _{ij} - (H^kW^k)_{ij})^2}\)\\
             $O^{k+1} \leftarrow F(W^k)$\\
            }
         }
     \caption{Traffic partitioning}
    \label{alg:nmf}
\end{algorithm}

\begin{figure}[!ht]
\begin{center}
\leavevmode
\includegraphics[width=1\columnwidth]{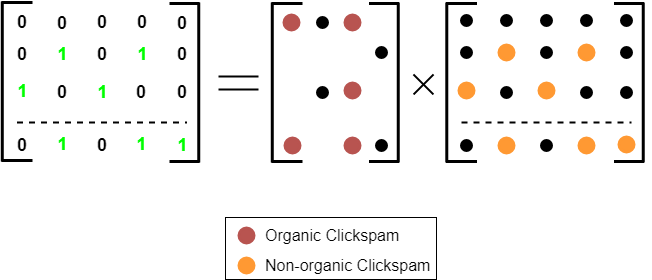}
\end{center}
\caption{Clicktok traffic partitioning}
\label{fig:owh}
\end{figure}

\subsubsection{Uniqueness of click partitioning}

In order to detect clickspam, the partitioning process uses
multiplicative rules to drive the optimisation function. We now show
that the multiplicative update rules used in Algorithm~\ref{alg:nmf}
lead to a unique optimal solution.

In each layer of our algorithm, as per Equation~\ref{eqn:nmf},
 $O$ is partitioned into $H$ and $W$. The probability of observing $t$
clicks given $r$ timing-patterns $H_1,\dots,H_r$ is given by
\(P(O_{ij}=t | H_1,\dots,H_r) = \prod_{k=1}^r P(O_{ij}=t | H_k)
P(H_k)\), where $P(H_k)$ is the prior probability about pattern $H_k$
(which can be set using a honey pot or drawn from a uniform
distribution). We'll ignore the layer number in the discussion that
follows.


Assuming that legitimate clicks can occur at any point in time, on
a per-source basis, the interclick arrival times between subsequent
clicks justifies a Poisson distribution. The average number of
legitimate clicks in any time interval can be well approximated by a
normal distribution~\cite{vjethava:billingsley95}. Making no
assumptions about the distribution of clickspam, the probability of
observing $t$ legitimate clicks due to pattern $H_{*k}$ in the
$i^{th}$ user at the $j^{th}$ time interval is given by the Poisson
probability distribution: \(P(t, i, j, H_k) = \frac{\lambda_{i,j,k}^t \exp(-\lambda_{i,j,k}) }{t!} \). After computing the log-likelihood, we obtain:
\begin{equation}
  \ell = \sum_i^n \left(O_{i*} \ln(H^TW_{*i}) - H^TW_{*i} - \ln(O_{i*}!)     \right)
\label{eqn:opt}
\end{equation}


The observations in matrix $O$ are the number of ad-clicks (per time
interval) from multiple simultaneously active (hence superimposed)
basic-patterns. Considering all the patterns $H_{*1},\dots,H_{*r}$,
let $H_{kj}$ be the probability that pattern $H_{*k}$ has been
exploited to generate clicks in time interval $j$. The intensity of
click fraud contribution of $H_k$ to traffic from user $i$ is given by
$W_{ik}$. The corresponding vector $W_{*k}$ encodes the commonality of
pattern $H_k$ across all users (\( \sum_k W_{ik}=1\) ).  Therefore
\(O_{ij} = \lambda_{i,j} = \sum_k^r H_{ik}W_{kj}\). Replacing $(HW)$
by $\lambda$ in Equation~\ref{eqn:opt} and computing derivatives with
respect to $\lambda$, we obtain:

%

\begin{eqnarray}
  \ell &=& \sum_t t \ln\lambda - \lambda - \ln(t!)\\
  -\frac{\mathrm{d}\ell}{\mathrm{d}\lambda} &=& \sum_t -\frac{t}{\lambda} + 1\\
  -\frac{\mathrm{d}^2\ell}{\mathrm{d}\lambda^2} &=& \sum_t \frac{t}{\sqrt(\lambda)} \label{eqn:seconddiff}
\end{eqnarray}

From Equation~\ref{eqn:seconddiff}, we see that
$-\frac{\mathrm{d}^2\ell}{\mathrm{d}\lambda^2}$ is positive for all
values of $\lambda$. Hence, our optimisation function is concave. This
means that there are no local maxima which will adversely affect
optimisation.  The solution is unique regardless of the initial values
of $W$ and $H$.


\subsubsection{Isolating clickspam}
\label{sec:validation}
This stage of the application of NMF algorithm is a standard
  step, which may be familiar to the reader experienced in NMF. Here we
  reverse the partitioning process, in order to arrive at the traffic
  of interest i.e. clickspam.

We isolate organic clickspam, by observing the patterns
 (in matrix $H$) that repeat (as given by the activation
matrix $W$). 
Each column of weight matrix $W_{j*}$ gives the extent of repetition
of the $j^{th}$ base pattern $H_{j*}$. All patterns that repeat twice
or more, are involved in organic click fraud. Using matrix $H$, we
compute $H'$ as follows; we retain all $H'_{j*}=H_{j*} \forall \sum
W_{j*} > 2$ and reset the rest $H'_{j*} = 0 \forall \sum W_{j*} \leq
2$ (i.e. we retain all traffic that corresponds to click fraud). The
weight matrix $W$ is also modified to retain weights corresponding to
click fraud $W'$, $W'_{j*}=W_{j*} \forall \sum W_{j*} > 2$ and reset
the rest $W'_{j*} = 0 \forall \sum W_{j*} \leq 2$. We then compute $O'
= W'H'$, activating click fraud patterns alone. Each cell of $O'$
contains the number of fake clicks detected during any 5 minute
interval.

Identification of non-organic click fraud is a two-step process. {\em
  First}, we cluster closely related pattern vectors using $k$NN ($k$
nearest-neighbors)~\cite{knn}, a centroid-based clustering
technique. The distance function between vectors is the inverse cosine
similarity function. {\em Second}, to determine which clusters of
patterns correspond to inorganic clickspam, we use entropy as the
validation metric. We compute the average Shannon {\em entropy} of the
distribution over inter-click times within click patterns in each
cluster. The use of statistical average is appropriate as the patterns
within a cluster are expected to be fairly similar to each
other. Together, the weight and entropy of a cluster can be used to
isolate different click fraud attacks. All clusters with entropy
greater the $0.5$ are indicative of legitimate traffic (which have
innately higher entropy), the remaining clusters consists of low-rate
clickspam generated using randomized or constant time offsets.

\subsubsection{Algorithmic complexity}
The complexity of our inference algorithm primarily depends on the
number of time intervals and the density of clicks in each
interval. There are $O(nmr)$ update operations in the worst
(dense) case in the first iteration.  We assume that the number of
legitimate clicks is roughly $log(m)$, based on the sparseness of user
click activity where $m$ is the number of time intervals. $r$ -- the
number of independent sources --- is a small constant. Thus the
complexity reduces to $O(n\ log\ m)$. The $k$NN algorithm used to
cluster the basic-patterns involves a total of $n log(n)$ comparisons,
where $k$ is a constant. Hence its complexity is $O(n log n)$.

\section{Evaluation}
\label{sec:eval}

In order to evaluate our inference algorithm, we
require access to click traffic which
  contain both organic and non-organic clickspam. First, we
  acquired a pre-labelled dataset, consisting of both
legitimate clicks and clickspam, in controlled
proportions. To achieve this, we collected
traffic within a university network, filtered it, and then
exposed it to a testbed; consisting of malicious apps and
click malware. As a result, the traffic exiting
  the testbed contains clicks from both sources of legitimate and fake
  clicks.

\subsection{Dataset Acquisition}
\label{sec:eval:dataset}

To collect {\bf legitimate ad-click traces}, we setup
traffic monitors on backbone routers of a university campus
network. For each click, an advertisement is requested and the traffic
monitors record the following information from the request: ad URL, ad
server IP address, publisher page (referrer URL), source IP address,
User-Agent string (UAstring) and the timestamp. Overall, we recorded a
total of $217$,$334$,$190$ unique clicks, between June 2015 and
November 2017. The data was collected after due process of obtaining
ethical approval, and all stored data is encrypted. 

\subsection{Click Malware and Exposure to Legitimate Traces}
Our malware dataset consists of $12,518$ binaries
  (allegedly) associated with clickfraud from a private collection. To
  verify that the malware dataset is indeed linked to clickfraud, we
  set up a dynamic analysis environment. The environment is suitably
  instrumented to study the dynamic behaviour of malware in response
  to legitimate clicks.

For Windows binaries, our analysis environment consists
of Linux Mint 17.3 servers. Each server includes 4 x AMD Opteron 6376
Sixteen-Core 2.30GHz CPUs equipped with 1TB RAM and can run several
VMs. Each VM consists of a Windows 8.1 guest OS installed with
instrumentation tools. We used the Selenium IDE 3.1~\cite{selenium}, to
inject click traffic into a Firefox 54.0 browser within the
VM. Outgoing click traffic from the VM is captured using Virtualbox's
network tracing facility. Dynamic analysis is carried out via the
following workflow:
\begin{enumerate}
\item Each windows binary is retrieved from the malware dataset and installed
  on a VM.
\item A fraction of click sources are chosen from the legitimate traces.
\item For each source (UAstring), a firefox browser is launched within
  a fresh VM.
\item For each source, the webpage hosting the ad (referrer URL) is
  pushed into the browser and the corresponding (legitimate) ad-clicks
  are injected into the browser, whilst preserving inter-click times.
\item The installed malware is thus exposed to legitimate click
  traffic. Any ad-clicks induced by malware are recorded on the VM and
  labeled as clickspam, on the basis that the ad-click was induced by malware.
\end{enumerate}

For Android binaries, our analysis environment is based on
  the same hardware, however the Virtualbox VM is configured to use an
  Android 6.0 guest image. Our analysis-workflow schedules an app to
  run on the VM using the Monkeyrunner tool~\cite{mrunner}. This tool
  allows us to inject ad-clicks from our dataset on to the installed
  app. We used a network monitoring tool to capture mobile clickspam
  locally on the device. It leverages Android's VPN API to redirect
  the device's network traffic through a localhost service enabling
  packet inspection. For Android apps, the workflow consists of the
  following:
\begin{enumerate}
\item Each android app is retrieved from the dataset and pushed on to
  a fresh VM where it is executed for a chosen duration of time.
\item As before, a fraction of click sources are chosen from legitimate traces and injected into the app using the Monkeyrunner tool.
\item If the app is inclined towards click fraud, we expect to see
  additional clicks and ad-fetches over and above those
  injected. These are recorded by monitoring outgoing network traffic
  using Lumen, and labeled as clickspam.
\end{enumerate}

Out of the $12,271$ Windows malware that were initially
  associated with clickspam, we found that only $9,773$ produced
  clickspam when put through our dynamic analysis
  environment. Similarly, out of the $247$ apps we inspected, $93$
  produced clickspam. This means that our traffic dataset is based
  on a total of just over 10k click malware samples.

\subsection{Passive Detection -- Mimicry}
\label{sec:eval:passive}

In order to understand the significance of contextual parameters, we
examine the traffic from {\em multiple ad networks}, control for
the effects of the {\em sizes} of legitimate and fake clicks, and
{\em multiple ad categories}. Furthermore, to evaluate the
performance of Clicktok as a passive defence, we are concerned
with the false-positive (FPR) and true-positive (TPR) rates. The FP
and TP rates are the fraction of legitimate clicks reported as
fraudulent clicks, and the fraction of fraudulent clicks detected,
respectively.

The first step is to create the traffic matrix, as explained in
Section~\ref{sec:inference:algdesign}. The data from each source IP
address is loaded into one row of the matrix, i.e. one row per source
per day. The input input series is then divided into five minute
intervals, thus establishing $288$ columns per day. For each interval,
we compute the total number of clicks; for which we do not know if the
click is a legitimate or spam click.

\paragraphb{Time interval (bin size):} We must next consider the level of temporal granularity required in isolating fake clicks.
Specifically, is the defender happy to know that a fake click
occurred within a specific day, hour or five minute interval. Within the limitations of our study, we discovered that the average
user clicks on less than 15 advertisements per day, and thus having a
small bin size in the range of a few seconds would be excessive in
terms of unduly high granularity.

\paragraphb{Number of basic patterns:} The number of unknown basic
patterns, $r$, is fixed by hand, and thus we
need to consider the maximum possible number of basic patterns,
$r=n$. consideration is a scenario, where the traffic
dataset is not compressible at all and therefore, a simple
way to choose $r$ is to simply set it to $n$. Subsequently,
  a higher value for $r$ showed no
resulting impact on detection efficiency, but
simply causes some basic patterns to repeat and cluster
together when $k$-nearest neighbours is
applied. An $r$ value greater than $73$ did not
result in any new basic patterns.

 %

\paragraphb{Attack volumes:} For evaluating our algorithm, we must also
consider the volumes of click fraud attacks. The hardest case for detection
involves stealthy click fraud attacks, which our dataset contains, where
the level of click fraud is less than 5 clicks per source device, per day
(e.g. as induced by the TDSS and TDL-4 botnets~\cite{chen2017measuring}). As well as stealthy click
fraud attacks, our dataset also consists of {\em sparse} click fraud attacks,
which are mid-range attacks that correspond to between 5 and 15 clicks per
day (e.g. Chameleon~\cite{kirubavathi2014botnets} and Zeus~\cite{binsalleeh2010analysis}
descendants). Due to the minimal amount of
clicks, any threshold-based statistical defence technique will find it
difficult to detect a stealthy attack. Finally, we also consider
{\em firehose attacks}, which involve attacks with volumes greater than
15 clicks per day, per source (e.g. ZeroAccess~\cite{zeroaccess}). The attack traffic in
this case, is distributed amongst publisher sites, to reduce the
per-publisher volume below the anomaly threshold~\cite{metwally:icdcs07}.

\paragraphb{Isolating organic clickspam:} When a subset of legitimate
clicks is reused, even when click times are partially
  randomised or multiple legitimate clickstreams are combined, it
triggers the optimization criteria within the inference algorithm. The
organic click traffic used in the click fraud attack is
  found in matrix $H$, while the clickspam is identified by the
location of pattern activation in matrix $W$ (weight
  matrix). The precise mechanism is given in
Section~\ref{sec:validation}.

\paragraphb{Isolating non-organic clickspam:} For isolating
  non-organic clickspam, we first set the cosine-similarity threshold
  to $0.9$ over non-repeating column vectors of the set of basic
  patterns $H$. We discovered several giant clusters
(between 1 and 25) that contained a majority of the clicks from the
clickstream ($70\%$ to $98\%$). As well as this,
  we identified smaller clusters (between 0 and 16) which contained
  fewer clicks (between $2\%$ and $30\%$ of the clicks). 
  From these observations, we then examined the normalized entropy
and weight of each cluster of patterns. As a heuristic, clusters of
low entropy ($\leq 0.5$) correspond to simple click fraud attacks
and pattern clusters with a normalised entropy greater than
  $0.5$ are considered to be legitimate flows. The clusters of low
  entropy consist of attacks where the click modules are characterised
  by low-variability or near-constant inter-click times. This includes
  click modules that generate traffic via constant or random offsets
  to legitimate clicks.

\begin{table}
\centering
\scriptsize
\begin{tabular}{@{\extracolsep{0em}}llccc@{}}\hline
ad network (duration) & Attack  & \#spam/src/day  & \% FPR    & \% TPR \\\hline
Google (1 week) & stealth-1  &  1--4       &  0.066   &  62.80\\
               & sparse-1   &   5--15      &  0.009  &  74.31\\
               & firehose-1 &   >15    &  0.004  &  87.46\\

Google (12 weeks) & stealth-12 &    1--4    & 0.019   & 78.03\\
               &   sparse-12   &    5--14   & 0.006   & 81.33\\
               &   firehose-12 &    >15 & 0.004   & 99.32\\
\\
adCentre (1 week)  & stealth-1 &  1--4    &  0.071    &  67.05\\%
                   & sparse-1  &  5--15   &  0.008    &  74.52\\
                   & firehose-1&  >15 &  0.004    &  81.07\\

adCentre (12 weeks) &stealth-12  &   1--4    &  0.024   &  81.79\\
                    &sparse-12   &   5--15   &  0.005   &  82.92\\
                    &firehose-12 &   >15 &  0.003    &  98.36\\\hline
\end{tabular}
\caption{Passive detection --- detection and error rates of inference}
\label{tab:multiplevol}
\end{table}
\normalsize

\paragraphb{Detection and error rates of inference:}
The results of applying Clicktok are summarized in Table~\ref{tab:multiplevol}. We observed fairly serviceable detection rates, between 70\% to 100\%. More importantly, the false-positive rates are fairly low, with rates between
3 clicks to 66 clicks per hundred thousand clicks, for high-volume and
stealthier low-volume attacks respectively. These results are consistent across ad networks, which aids with verifying the evaluation results of Clicktok in a realistic setting.


\paragraphb{Size of background traffic:} Upon evaluation of Clicktok, we
observed that larger traffic sizes improve inference. For the Google
ad network, the detection rate of stealthy attacks improved by 12\%,
whilst the FPR reduce from 66 to 19 clicks per hundred thousand.
Interestingly, we also observed this with adSense (Microsoft), which
increases confidence in the result. For stealthy attacks, this
significant reduction is achieved by exploiting correlations across
user clickstreams. As we consider the traffic for multiple users together,
even a low attack rate can be detected across click traffic from multiple
users. Understandably, we observed a lesser improvement with the firehose
attacks, as attack rates already afford better detection rates and lower
false-positive rates, even with just a few days of traffic.

\begin{table}
  \scriptsize
  \subfloat[Google 84M]{
      \begin{tabular}{@{\extracolsep{0em}}lccc@{}}\hline
        Fraud-type & \#spam & \% FPR    & \% TPR \\\hline
        Sponsored  & 16795     &  0.005\% &  93.595 \\
                   & 138345    &  0.005\% &  95.005\\
                   & 1332910   &  0.004\% &  95.692\\
\\
       Contextual &  22394 &  0.005\%    &  87.806 \\
                  & 171883 &   0.005\%   &  89.202 \\
                  & 1818777 &  0.004\%    & 90.546 \\
\\
        Mobile    &  18475  &  0.004\%    &   91.379\\
                  & 108999  &  0.003\%    &   92.833\\
                  & 1165221 &  0.003\%    &   92.654\\ \hline
  \end{tabular}}
  \hspace{10pt}
  \subfloat[Microsoft adSense 78M clicks]{
    \begin{tabular}{@{\extracolsep{0em}}lccc@{}}\hline
Fraud-type & \#spam & \% FPR    & \% TPR \\\hline
Sponsored & 18219   &  0.005\%  & 89.11\\
          & 123442  &  0.004\%  & 90.70\\
          & 912480  &  0.004\%  & 92.17\\
\\
Contextual & 20380   &  0.006\%   &   88.30\\
           & 323302  &  0.004\%   &   91.68\\
           & 2198249 &  0.004\%   &   90.93\\
\\
Mobile     & 10594   & 0.005\%    &   90.33\\
           & 141077  & 0.003\%    &   91.52\\
           & 1161338 & 0.003\%    &   94.76\\ \hline
    \end{tabular}}
    \caption{Detection and error rates of Inference across multiple clickstreams}
\label{tab:multiple}
\end{table}
\normalsize

\paragraphb{Effects of click category:} We examined click
  fraud in three categories: sponsored, contextual and mobile
  ads. Sponsored search ads are advertisements displayed by search
  engines, based on the keywords within a user's query alongside
  search results.  Contextual ads are a more generic form of
  advertisements, which are displayed on a webpage based on the
  keywords present on that webpage.  For instance, an advertisement on
  purchasing gold bars may be displayed on a webpage that contains
  information about investing in gold. Finally, mobile ads is a
  category of advertisements which are exclusively displayed on mobile
  devices. In all cases, the detection rates are fairly high, as well
  as a serviceably low false-positive rate. The detection of mobile ad
  click fraud has a slightly lower FPR compared to other
  categories.

\subsection{Active Detection -- Bait Clicks}
\label{sec:eval:active}
The idea of the active defence is that the ad network injects {\em
  bait clicks} with a well defined inter-click delay pattern along with the ad it serves. Any
mimicking of the injected pattern is evidence of click
fraud. This approach is loosely motivated from traffic-analysis
literature~\cite{RSG98}. We used an injected pattern where consecutive
injected clicks are a $\delta_n=\delta$ time apart $ t_n = \delta_n +
t_{n-1}$. The ad network can use more than one injected patterns each
defined by a different $\delta$ in order to keep the bait-clicks
discrete.

To implement this defence, as before the traffic matrix $O$ is
initialized with each row containing the time series from a 24 hour
output of the testbed. We then decompose $O$ using our inference
algorithm and $r=n$, however there are two important changes. First,
the initial $r$ rows of $H$ are set to the injected patterns instead
of being initialized with random value. Second, instead of applying
the update functions for $H$ from Algorithm~\ref{alg:nmf} for
all $i < r'$, we fix the values of $H_{i*}$ for all $i \leq r$, i.e. the
rows of $H$ representing injected patterns are not altered. This step
allows the other rows of $H$ to be suitably altered so as to represent
legitimate timing patterns. The final step is the isolation of fraud
clicks. This is done by analyzing $W$.  $W_{ij}$ gives the influence
of pattern $H_{j*}$ on click traffic time series $O_{i*}$. Since the
injected timing patterns are located on the first $r$ rows of $H$, the
fraction of fraud clicks for time-window $i$ of click traffic is
simply computed as: \( \frac{\sum^r_{j=1}
  W_{ij}}{\sum^n_{l=1}W_{il}}\).

In engineering terms, an ad network sends ads
encapsulated with JavaScript code into the user's browser. Code
execution is triggered by a suitable JavaScript event (such as when a
page has finished loading). For the ad network the network and
computational overheads are constant time per user and scales linearly
with the number of users. The impact on the user device is also fairly
minimal, dispatching three-four mouse clicks on an advertisement.

\begin{table}
\centering
\scriptsize
\begin{tabular}{@{\extracolsep{0em}}llccc@{}}\hline
ad network (duration) & Attack  &  \#spam/src/day & \% FPR    & \% TPR \\\hline
Google (1 week)& stealth-1  &   1--4      &  0.051  &  66.40\\
               & sparse-1   &   5--15     &  0.010  &  78.61\\
               & firehose-1 & >15     &  0.004  &  93.48\\

Google (12 weeks) & stealth-12 &    1--4       & 0.004   & 89.34\\
               &   sparse-12   &    5--15      & 0.004   & 91.62\\
               &   firehose-12 &    >15    & 0.003   & 96.77\\
\\
Microsoft (1 week)  & stealth-1 &  1--4     &  0.060    & 51.02\\%
                   & sparse-1  &  5--15     &  0.003     & 75.14\\
                   & firehose-1&  >15   &  0.005    &  92.60\\

Microsoft (12 weeks) &stealth-12  &   1--4      & 0.004   &  90.78\\
                    &sparse-12   &   5--15      &  0.003   &  92.44\\
                    &firehose-12 &   >15    &  0.002    &  95.41\\\hline
\end{tabular}
\caption{Active defence --- detection and error rates of inference}
\label{tab:activedefences}
\end{table}
\normalsize

\paragraphb{Results:} Active defence improves detection rates by
almost 10\%. The results of active defences are documented in
Table~\ref{tab:activedefences}. In both Google and AdSense, active
defences are very successful ($>89\%$) at detecting fake clicks at all
ranges of attack traffic volumes from stealthy to a firehose, at low
FPR of 30--40 per million clicks. The reduction in FPR for low-rate
attacks is most improved compared to passive defences, indicating the
importance of considering active attack approaches in fighting
click fraud.

However, when active defences are presented with poor context (1-week
traffic set), i.e. applied over only few clicks per user, we observe
that detection and FPR are similar to passive
defences. To understand why, we must note that looking for a response
to injected traffic mainly detects mimicking attacks (as opposed to
other variable-rate attacks such as randomly generated
fake clicks). In our dataset, a week's traffic contains between 2 and
43 user clicks per day. At attack rates of two per day, Clicktok fails
to detect the attack. As attack rates which are still fairly
stealthily increases, click fraud attacks are readily detectable; For
an increase in fake click rate from 1\% to 10\% of legitimate traffic,
we observe a reduction in FPR by an order of magnitude for stealth
attacks, while detection rate increases from 50\% to ~70\%. For higher
attack rates, the click fraud campaigns are mostly randomly generated
fake clicks, and these result in modest changes in FPR.

\section{Discussion}
Successful click fraud campaigns not only need to {\em scale}, they
must also be {\em credible (pseudo-legitimate)}, and {\em
  stealthy}. Attackers achieve scale using compromised apps to
automate click fraud campaigns and achieve stealth by using organic
clicks to institute click fraud. To combat stealth, Clicktok detects
click fraud based on timing characteristics of click traffic feeds
received at ad networks. Unlike threshold based defences, it exploits
correlations in timing behaviour across multiple compromised end-user
devices.

Detecting click fraud using timing information exploits a fundamental
property of attacks based on organic click fraud --- some attack
traffic is a function of historical legitimate traffic. Clicktok's
strength lies in its generic approach. Instead of assuming specific
attack methods for click fraud attack strategies, Clicktok uses a
notion of a {\em compressive} optimization function to isolate clicks
generated by different sources without prior knowledge about what
those sources are. Thus Clicktok can isolate non-organic clickspam
such as those generated at pseudo-random times and the replay of
organic clickstreams.

Our techniques show some promise: passive defences have a detection
rate of around 78\%--81\% with a false-positive rate of 190 per
million clicks; and, active attacks work best, with the false positive
rate entering a region of 40 to 70 false positives per million
clicks. This means the ad network charges the advertiser for 40 to 70
clicks per million clicks received, as compared with a false-positive
rate of between 200,000--300,000 clicks per million in current
ad networks~\cite{bluff,dave:sigcomm:2012}.

\paragraphb{Passive defences:} When Clicktok is applied as a passive
defence, it acts as a (lossy) compression function that partitions the
input time series into basis patterns, such that all basis patterns
from the same distribution are grouped together. Thus given a
clickstream containing clicks generated from different statistical
distributions (random variables) over generation times, Clicktok
creates as many partitions.

\paragraphb{Active defences:} We model a clickstream as a timing
channel between a user and the ad network composed of inter-click
times between consecutive ad clicks. The ad network watermarks the
channel periodically using bait-clicks. When an attacker harvests and
reuses a (legitimate) clickstream, it sets off watermark detectors
located in the ad network. Unlike conventional watermarks however, the
timing pattern induced by click fraud may not appear as an isolated
series of consecutive clicks. Clickstreams arriving at the ad network may therefore
contain clickspam that's thoroughly mixed (superimposed) with
legitimate clicks. Thus, the challenge of designing the watermark
detector system is to {\em unmix} the stream back into legitimate and
clickspam. Clicktok develops the algorithmic basis for carrying out
this work. In order to bypass the active defences, the attacker must
distinguish between the bait-click watermark and legitimate traffic,
which roughly speaking, shifts the burden of solving the click fraud
problem to the attacker.

\subsection{Limitations of Clicktok}

\paragraphb{Metrics used:} We use entropy as a validation metric with
a normalised threshold of $0.5$. However, this may be vulnerable to
the adversary generating non-organic clickspam using pseudo-random
number generator that generates clicks from distributions with high
entropy. While we haven't observed this taking place in our testbed
across 1.5 years, it is a reasonable countermeasure for the
attacker. Clicktok can still function in these circumstances by
leveraging a honeypot to populate the pattern matrix $H$ as follows:
columns of the pattern matrix are pre-populated with traffic from the
honeypot and fixed, i.e. excluded from the application of multiplicative
rules in Algorithm~\ref{alg:nmf} while the rest of the columns are
optimised. Upon convergence, the weight matrix will contain high
weights if any traffic similar to the honeypot is observed in the
traffic matrix $O$.

\paragraphb{IP aggregation and churn:} Often, enterprise networks may
deploy DMZ or other traffic aggregators to avoid exposing IP addresses
to the outside world. This impacts the extent of attribution. Malice
will therefore be at best traced back to the aggregator and further
investigation would be required to isolate the actual source behind
the aggregator. Churn causes similar record-keeping problems. In order
to positively attribute malice to a source, the source-IP addresses
involved must be reconciled with the local DHCP records. We note that
the impact on detection itself is minimal; an short DHCP expiration
policy, simply implicates both the previous and new source IP
addresses of a malicious source.

\paragraphb{Cookies and deletion:} A relatively reliable approach is
to use cookies instead of source-IP addresses to keep track of
malice. ad networks can track clickstreams on a per user basis using
authenticated HTTP sessions, instead of solely depending on source IP
addresses. Cookies can help address churn issues where a pool of
source IP addresses are cycled among many users.

\subsection{Comparison with related work}
We compared Clicktok's efficiency with click fraud detection techniques
 proposed in the literature. While some of these techniques
were not intended to work with the challenges of embedded fake clicks
or stealthy click fraud attacks, we can nonetheless compare against
them on real datasets to get a sense of how Clicktok compares against
these approaches. Several techniques to differentiate fake clicks from
legitimate clicks have been proposed.  Work in this space can be
categorized into {\em threshold-based} techniques that build a
baseline of benign behaviour and analyze deviations from the baseline,
and on timing analysis techniques. We choose two methods from timing
analysis and a recently proposed threshold technique for detection
efficiency.

\begin{table}
\centering \scriptsize
\begin{tabular}{@{\extracolsep{1em}}llccc@{}}\hline
Technique  & Attack  &  \#spam/src/day & \% FPR    & \% TPR \\\hline
Clicktok       & stealth  &   1--4     &  0.066  &  62.80\\
               & sparse   &   5--15    &  0.01   &  74.31\\
               & firehose &   >15  &  0.004  &  87.46\\

Similarity Seeker & stealth &    1--4    & 14.41   & 57.49\\
               &   sparse   &    5--15   &  9.68   & 59.82\\
               &   firehose &    >15 &  0.78   & 85.21\\

ViceROI        & stealth  &  1--4      &  10.23    &  60.03\\%
               & sparse   &  5--15     &   2.65    &  66.13\\
               & firehose &  >15   &   0.5     &  78.29\\

PubCrawl       &stealth  &   1--4       &  4.70    &  52.64\\
               &sparse   &   5--15      &  3.24    &  67.28\\
               &firehose &   >15    &  0.85    &  77.91\\\hline
\end{tabular}
\caption{Comparative analysis of Clicktok (Passive) vs others}
\label{tab:comp}
\end{table}
\normalsize

\begin{itemize}
\item Threshold-based approaches detect hotspots of activity between click-malware and publishers. One set of techniques detect traffic hotspots~\cite{metwally:icdcs07,metwally:www07,metwally:vldb08}. Another technique is to examine publisher-user pairs with above-average click rates~\cite{dave:ccs:2013}. All the techniques in this approach develop a normative baseline of activity and detect malicious behaviour beyond a threshold distance from the baseline. The idea is that fraudsters need to scale their activity to a level where their turnover (from a click fraud campaign) covers their costs as well as generate a  profit.

\item In time-series analysis, techniques such as auto-correlation, partial-correlation, and cross-correlation techniques~\cite{wei:1994:time} are used to process the input clickstream. Each clickstream is converted into a time-series by counting the number of clicks within each time interval, just as in Clicktok. Parsing Clicktok's traffic matrix in a row-major fashion gives the required time-series vector over which analysis techniques are applied. Recently, PubCrawl~\cite{pubcrawl}  extended this idea by combining it with learning approaches, where a classifier is trained with a labelled dataset containing time series for both honest and fraud clicksets.
\end{itemize}

The algorithmic complexity of both is similar to Clicktok so we were
able to readily run them on the 1-week Google ad network
dataset. Results from our comparison are shown in
Table~\ref{tab:comp}. We found that the performance of Clicktok is similar
to existing solutions for high-rate (hose and firehose) attacks in
terms of detection rate, but has a much better FPR for all attacks. For
low-rate attacks (stealth and sparse), Clicktok significantly
outperforms all existing solutions. For example, for a hose attack,
Clicktok's FPR is 0.04\% whereas FPR for other approaches ranged from
0.5\%---7.65\%. As another example, for a stealth attack, Clicktok's
FPR is 0.066\% whereas FPR for other approaches ranged from
4.7\%--14.7\%.

Our experiments show the limitations of using threshold based
approaches, which can be defeated by reducing the network loads placed
by attackers. In the case of ViceROI, the cost of renting a botnet has
fallen by three orders of magnitude which allows attackers to scale up
the number of attack hosts for the same amount of click fraud, pushing
the level of fraud per user and per publisher below the detection
threshold. Similarly, with Similarity-Seeker which looks for traffic
hotspots, but without using bluff ads to generate a baseline, the
hotspot is diffused with increase in the number of attack hosts and
malicious publishers. Interestingly, PubCrawl was fairly successful at
detecting stealthier attacks than Similarity Seeker and
ViceROI. PubCrawl also uses time-series analysis, however it is not
designed to handle the case where clickspam and legitimate traffic are
temporally overlapped. This capability gap is especially evident when
the fake clickstream size is smaller as any overlap tends to severely
'damage' the signal that PubCrawl detects.

\section{Related Work}
\label{sec:relwk}

Click fraud demonstrates a serious economic impact on the Internet
sub-economy of the click marketplace, bringing rise to a growing
body of academic research to address the problem. Click fraud inflicts
losses for tens of thousands of online advertisers, causing upwards
of hundreds of millions of dollars each year~\cite{mungamuru:fc08}.

Several papers have highlighted the importance of the field of
advertising and
click fraud~\cite{Stone-Gross:2011:UFA,zarras:2014:imc}.  Keeping
ad networks free of fraud is highlighted by the work of Mungamuru et
al.~\cite{mungamuru:fc08}, who show that ad networks free of
click fraud results in a competitive advantage over rival ad networks
and thus attracts more advertisers.  Research shows that even the
largest advertising platforms are affected by click
fraud~\cite{kitts2015click} and are tackling the problem, by primarily
employing data mining techniques to distinguish legitimate, fraudulent
or bot-generated click events.  Clickbots are a leading attack vector
for carrying out click fraud; around 30\% of the clicks are fraudulent
across major ad networks~\cite{dave:sigcomm:2012,bluff} and originate
from malware networks (rent-a-botnet services).

\paragraphb{Clickbots and Malware Networks:} Malware networks are
primarily operated by malicious publishers or a malicious advertiser who depleted the budget of competing advertisers such as the WOW botnet~\cite{lam:09:wow}.
Chen et al.~\cite{chen2017measuring} discovered that the
TDSS/TDL4 botnet --- one of the most sophisticated botnets ---
incurred an average of \$340 thousand in daily losses to
advertisers. The losses scale up to ten times more than the daily
impact some previous botnets had to the advertising
ecosystem~\cite{pearce2014characterizing,meng2013dns}.  Naturally, a
number of studies have focused on botnets used for click fraud. Daswani
et al.~\cite{daswani:hotbots:2007} reverse engineered clickbot and
followed by Miller et al.~\cite{miller:dimva:2011} on Fiesta and
7cy. These works focused on the C\&C traffic structure of the attack
infrastructure. We consider the timing characteristics of
click-generation algorithms used by botnets instead of the structure
of the C\&C traffic or the malware binaries. Unlike these specialized
studies which are solely focused on specific botnets, our work has
wider application including future botnets.

Online advertisement networks employ a variety of heuristics to detect
click fraud and apply corresponding discounts on advertisers they
invoice~\cite{clickspamacct}. These heuristics involves tracking and
bounding user click behaviour often on a per-IP-address basis. In
response, criminals are using large distributed attack
networks of bots for launching replay click fraud attacks. Bots are unique
in that they collectively maintain a communication structure across
infected machines to resiliently distribute commands from a {\em
  command and control} node. The ability to coordinate and upload
click fraud attack commands gives the botnet owner vastly
increased power against major ad networks.

\paragraphb{Traffic analysis:} Several
countermeasures~\cite{metwally:icdcs07,metwally:www07,metwally:vldb08}
are based on traffic analysis. These methods isolate machines being
used for click fraud attacks by identifying traffic hot-spots between
the attack machines and publisher sites. Not only non-stationary
traffic, but also stealth occurrences of malicious behaviors both
introduce issues to analyse anomalous network
traffic~\cite{khanchi2018streaming}.  Machine learning frameworks have
been used to help analyse network behaviour, identify click fraud and
adapt to changes in traffic
~\cite{khanchi2018streaming,mouawi2018towards}. However, most of of
this research is done on botnet-induced click traffic, and does not
identify the effectiveness of these methods on other forms of
fraudulent clicks.

\paragraphb{Detecting rogue publishers:} In~\cite{dave:ccs:2013}, Dave
et al. point out that the conversion rate of malicious publishers is
abnormally high. Thus a heuristic based on the conversion rate could
be used as part of a click fraud defence. However, this is easily
circumvented by distributing attack traces across a few publisher
accounts. Or, alternately after an account is ``tainted'', it is
abandoned. This strategy imposes additional effort on the attackers
(they need to register new publisher accounts with high frequency) but
this is cheap. Such a strategy is already well in use to evade
reputation blacklists for domain names~\cite{grier:12:eaas} due to
which the mean life-time of domains is a mere 2.5 hours.

\paragraphb{Click honeypots:} Haddadi proposed the use of Bluff
ads~\cite{bluff} -- a fake ad that is of little interest to the
legitimate user but would attract click fraud attacks in the
wild. Bluff ads can be used to collect attack traffic and in
conjunction with Click fraud to extract basis patterns corresponding
to the static timing characteristics of malware click-modules. Instead
of bait-ads, Clicktok applies bait clicks to entice click fraud apps
into responding with watermarked click fraud. This is a new approach
towards instrumentation of click fraud apps at scale, and will form
part of our future work.


%
\paragraphb{Human bots:} A human-centric approach to setting up a
call-centre to generated handcrafted
clickspam~\cite{kshetri:ieeesnp10} and tricking people into clicking
ads on porn websites and mobile gaming
apps~\cite{felt:survey11}. Zhang et al. studied click fraud by
purchasing click traffic feeds. They also showed that the explicit
existence of human clickers~\cite{zhang:webquality:2011}. Our work is
focused on clickspam generation from malware alone.

\paragraphb{Trustworthy clicks:} Gummadi et al.~\cite{gummadi:nsdi:2009}
propose to combat bot activity through detection mechanisms at the
client. The client machine has a trusted component that monitors
keyboard and mouse input to attest to the legitimacy of individual
requests to remote parties. In addition, Juels et
al.~\cite{juels:usenixsec07} likewise propose dealing with click fraud
by certifying some clicks as premium or legitimate using an attester
instead of attempting to filter fraudulent clicks.
Zingirian and Benini~\cite{DBLP:journals/corr/abs-1802-02480} show
that a single adversary can increase the click count for given
advertised subscribers, from a single IP address, even if the
paid clicks are not linked with legitimate advertisements. They
evaluate security tradeoffs and revenue losses from discarded
clicks claimed as being illegitimate, and formulate an algorithm to
control the tradeoff; which shows to induce a very low impact even
with a large volume of clicks. Kintana et al.~\cite{kintana:sre:2009}
created a system designed to penetrate click fraud filters in order
to discover detection vulnerabilities. Blundo et
al.~\cite{blundo:seke02} propose to detect click fraud by
using Captchas and proof-of-work schemes. In a proof-of-work based
scheme~\cite{dave:sigcomm:2012}, the ad network serves Captchas
probabilistically in response to ad-clicks to verify the authenticity
of clicks. The main problem here is that malware often delegate the
problem of solving a CAPTCHAS to a third party. For example, via a
gaming app where the players are asked to periodically solve Captchas
for ``authentication purposes''.

\paragraphb{Detecting fake search engine queries:} A number of
click fraud malware use cover traffic in the form of automated search
engine queries leading to the malicious publisher's
website. Researchers have dedicated considerable effort to methods for
differentiating search queries from automated and human sources. Kang
et al.~\cite{kang:2010:www} propose a learning-based approach to
identify automated searches but exclude query timing. Yu et
al.~\cite{yu:wsdm:2010} observe details of bot behaviour in aggregate,
using the characteristics of the queries to identify bots. Buehrer et
al.~\cite{buehrer:2008:airweb} focus on bot-generated traffic and
click-through designed to influence page-rank. These efforts do not
examine timing behaviour, focusing instead on techniques for the
search engine to identify automated traffic. Shakiba
et al.\cite{shakiba2018spam} propose the detection of spam search
engine queries using a semi-supervised stream clustering method,
which characterises legitimate and illegitimate users using
linguistic properties of search queries and behavioural
characteristics of users; shown to be accurate 94\% of the time
with low overhead. Graepel et al.~\cite{graepel:icml2010} define a scalable Bayesian click-through rate prediction
algorithm, mapping input features to probabilities, for Sponsored Search
in the Bing search engine. They showed that their new algorithm was
superior to, and outperformed a calibrated Naive Bayes algorithm ---
regardless of the new algorithm being calibrated.

\paragraphb{Machine-learning based defences are vulnerable to mimicry
  attacks:} Many machine-learning based
approaches~\cite{Stone-Gross:2011:UFA} have been proposed for
click fraud detection. Many of these rely on rich feature sets. While
there are clear benefits to a diverse feature set we also know that
some features are highly predictable. For instance, accurately
predicting ads of interest to a user is
possible~\cite{graepel:icml2010}, reducing the utility of using
features such as keywords in detecting click fraud --- malware can
target advertisers relevant to the user. Our finding is that timing
information is relatively hard to predict and should be used in
conjunction with other features to combat click fraud.

\paragraphb{Conventional time-series analysis techniques don't work:}
Conventional time-series techniques are auto-correlation,
partial-correlation, and cross-correlation
techniques~\cite{wei:1994:time} which can find statistically similar
subsets across two or more time series. Correlation tools are typically
applied to wavelet coefficients or to the inverse transform of wavelet
coefficients at carefully selected coefficient
levels. Correlation-based techniques can detect sub-similar features
if the time series signal is not contaminated by convolutions of the
signal with other signals or high-amplitude noise. Unfortunately, in
the case of click fraud detection the traffic is contaminated by
time-overlapping legitimate and spam clicks as shown in
Figure~\ref{fig:inputpattern}. Indeed as prior art has noted, compromised-apps or click-malware generate clickspam only when some ``trigger''
fires such as the presence of legitimate clicks on the compromised
device~\cite{moser:07:multipath}.

\if 1
Fine-grained time-series analysis allows for collecting features which
aid in accurate fraud detection~\cite{oentaryo2014detecting}. Mobile
advertising data can be complex; involving heterogeneous information
and noisy patterns with missing (or invalid) data. Ensemble methods
show to offer promising results to classification tasks with mixed
types and noisy/missing patterns.

 Wavelet analysis~\cite{percival:2006:wavelet} partitions
time series into different frequency ranges with varying temporal
precision. This does not map into a separation of signal components by
the distribution used to generate them. Thus wavelets-based
partitioning can be a useful component of a wider algorithmic
framework as a noise removal technique. However, it cannot be used to
isolate click fraud on its own since it isn't designed to separate
superimposed click patterns.

\fi
Time series analysis has been used previously for detecting
click fraud. In PubCrawl~\cite{pubcrawl}, the main idea is that users
tend to be a lot more noisier than crawlers who have a relatively
stable behaviour pattern across time. The challenge we address is
significantly harder: what happens when machines (crawlers) mimic
user-behaviour by closely following past user behaviour rather than
the random strategies (which PubCrawl is designed to isolate).

\section{Conclusion}
\label{sec:conclusion}
Online advertising is a funding model used by millions of websites and
mobile apps. Criminals are increasingly targeting online
  advertising with special purpose attack tools called {\em click
    malware}. Click fraud instituted via malware is an important
  security challenge. Static attacks involving large attack volumes
  are easily detected by state-of-the-art techniques. However, dynamic
  attacks involving stealthy clickspam, that adapt to the behaviour of
  the device user, are poorly detected by current methods. We found
  that timing analysis can play a compelling role in isolating click
  fraud, instituted via both static and dynamic attack techniques. We
  applied NMF, a technique that identifies clickspam by exploiting the
  relative uncertainty between clickspam and legitimate
  clickstreams. It achieves this by detecting repetitive patterns that
  arise within ad network clickstreams from organic clickspam. We
  analysed a corpus of malware within an instrumented environment, that
  enabled us to control the generation of clickspam by exposing
  malware to legitimate clickstreams. We tested a passive technique
  which shows some promise. We also evaluated an active defense, where
  we injected watermarked click traffic into the analysis environment,
  that works better still. While timing analysis is well studied
  within the field of information hiding, for its ability to unearth
  hidden communication, its potential has yet to be fully explored in
  understanding stealthly click fraud attacks. Our work indicates that
  timing analysis might indeed be relevant to building better
  click fraud detection.

\bibliographystyle{ACM-Reference-Format}
\bibliography{paper}




\end{document}